\documentstyle[12pt,epsfig]{article}
\begin{document}
\baselineskip 3.9ex \def\be{\begin{equation}} \def\ee{\end{equation}}
\def\bea{\begin{eqnarray}} \def\eea{\end{eqnarray}}
\def\bd{\begin{displaymath}} \def\ed{\end{displaymath}}
\def\l#1{\label{eq:#1}} \def\eqn#1{(~\ref{eq:#1}~)} \def\no{\nonumber}
\def\av#1{{\langle  #1 \rangle}} \title{ Rooted Spiral Trees on
Hyper-cubic lattices\\
\vspace{0.2cm} Sumedha\thanks{sumedha@theory.tifr.res.in}\\
\vspace{0.2cm}  Department Of Theoretical Physics\\  Tata Institute Of
Fundamental Research\\ Homi Bhabha Road, Mumbai 400005\\ India}
\date{} \maketitle
\begin{abstract}
We study rooted spiral trees in $2,3$ and $4$ dimensions on a hyper
cubic  lattice using exact enumeration and Monte-Carlo techniques. On
the square  lattice, we also obtain exact lower bound of $1.93565$ on
the growth constant  $\lambda$. Series expansions give
$\theta=-1.3667\pm 0.0010$ and  $\nu = 0.6574\pm0.0010$ on a square
lattice. With Monte-Carlo simulations  we get the estimates as
$\theta=-1.364\pm0.010$, and $\nu = 0.656\pm0.010$.  These results are
numerical evidence against earlier proposed dimensional  reduction by
four in this problem. In dimensions higher than two, the  spiral
constraint can be implemented in two ways. In either case, our series
expansion results do not support the proposed dimensional reduction.

\end{abstract} 

Spiral structures are very common in nature. Some examples of the
beautiful spiral structures in galaxies, shoot arrangement in plants,
polymers with spiral structure etc may be found in the book by
Hargittai \cite{hargittai}. In statistical mechanics, lattice models
of spiral self  avoiding walks  have been studied and can be solved
exactly in two dimension \cite{privman,guttmann}, though no solution
is known for the self  avoiding walks without the spiral constraint. A
model of spiral trees  and animals was proposed by  Li and Zhou
\cite{li}, which based on  numerical studies was suggested to be in a
new universality class. This problem was further studied by Bose
et. al \cite{bose}.  Based on the numerical evidence, and guided by
the fact that magnetic field acting perpendicular to the motion of a
charged particle produces spiralling motion and reduction by two in
effective dimensionality, they conjectured that spiral tree problem
would show a dimensional  reduction by four. They conjectured the
exponents of the spiral tree  problem follow the following relations

\begin{eqnarray}
\theta = (d-4) \nu_{pl} ~ \mbox{for d = 2}\\   \theta = (d-4)
\nu_{\perp} ~\mbox{for d} > 2
\end{eqnarray}
where $\theta$ is the entropic exponent and $\nu_{pl}$ and
$\nu_{\perp}$  are the exponents related to the radius of gyration in
the plane in which  the tree has a rotational constraint and
perpendicular to that plane  respectively.

Since then this problem has not been studied further. Dimensional
reduction is an intriguing possibility. The lattice tree model without
spiral constraint is known to show a dimensional reduction by two
\cite{brydges}. The directed version, show a dimensional reduction by
one. For both models, the tree and animals are believed to lie in the
same universality class. In this paper, we revisit the problem and
obtain a significantly longer series for rooted spiral
trees. Specifically  in two dimensions we have added twelve terms to
the earlier series of $25$  terms. In three and four dimensions, we
generated a seventeen and a thirteen  term series respectively. The
earlier known series in three and four  dimensions had thirteen and
nine terms. In the process, we also correct  some mistakes in the
earlier reported series. We also perform Monte-Carlo (MC) simulations
using the improved incomplete enumeration algorithm  \cite{sumedha}
and generate spiral trees up to sizes of 1000 in two  dimensions. Our
analysis of exact series and MC samples do not support  the
conjectured dimensional reduction by four in this problem.

A lattice tree is a cluster of connected sites which contains no
loops. Spiral trees are a subclass of lattice trees. In a tree  every
cluster site is attached to the origin through a unique path.   In a
spiral tree, this path has a specific rotational constraint.

\begin{figure}
\begin{center}
		 \epsfig{figure=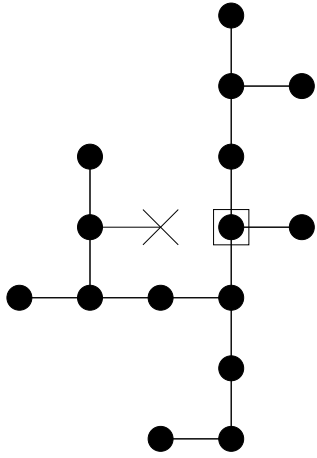,height=6cm}
\end{center}
\caption{A rooted spiral tree of 15 sites on a  square lattice. The
root is the site enclosed in the square. At the root site the tree has
freedom of choosing any of the four neighbouring sites. We count the
spiral tree by number of sites and hence all bonds between two
occupied sites is always assumed to be present. The site marked by
$X$,  if present will result in a loop for spiral site trees and hence
will not  be allowed. But it can be present in a  spiral bond tree.}
\label{sptr}
\end{figure}

We define rooted spiral tree as a acyclic connected subgraph of a
lattice such that the projection of the path joining any site of the
tree to the root on $x-y$ plane contains no left turn
(Fig. \ref{sptr}).  We will measure the size of a spiral tree by the
number of sites present  in the tree. These are called spiral site
trees.  The number of possibilities of spiral bond trees are more than
that for  spiral site trees but both are thought to lie in the same
universality  class. For example, the site marked as $X$ in Fig.1 is
not allowed in  spiral site tree as it would introduce a loop. But it
can be present in  a spiral bond tree.

Let the total number of  distinct rooted spiral trees be $A_n$.  This
is expected to have a asymptotic behaviour of the form

\begin{equation}
A_n \sim C {\lambda}^n n^{-\theta} \label{trno}
\end{equation}
where $C$ is a constant, $\theta$ is a critical exponent and is
expected  to depend only on the dimension of the lattice and $\lambda$
is known as  the lattice dependent growth constant. The existence of
growth constant  $\lambda$ for unrooted lattice trees and animals has
been proved rigorously  using concatenation and super multiplicity
arguments \cite{klein}. Also  a rigorous lower bound for $\theta$ for
unrooted lattice trees and animals  has been proved \cite{madras}
using pattern theorem. Specifically, it is  $\theta \ge (d-1)/d$, for
any dimension $d \ge 2$. Eq. \ref{trno} is expected to hold for most
cluster enumeration problems on regular lattices, though other
asymptotic forms are also possible. For example, for spiral
self-avoiding walks on square and triangular lattices, $A_n$ tends to
a stretched exponential in $n$ in the asymptotic limit
\cite{privman,guttmann}. Though we do not prove existence of $\lambda$
and $\theta$ for spiral trees in this paper, we derive a lower bound
for $\lambda$. Also, since spiral trees are a subset of lattice trees,
$\lambda_{spiral} \le \lambda_{all}$, where $\lambda_{all}$ is the
growth  constant for all trees, $\lambda_{all} \approx 3.795$ on a
square lattice \cite{jensen}. In two dimensions, we have derived the
generating function  for enumeration of a subset of all possible
spiral trees. The value of  growth constant for this subset is
$1.93565$. This gives a lower bound on  the growth constant
$\lambda_{spiral}$ of the spiral trees on a square  lattice. This
bound will be derived in Section 1.1.

For the conventional lattice animals, one can prove $\theta \geq 0$
through concatenation and super-multiplicity arguments \cite{klein,madras}. 
Concatenation does not work for spiral trees. Interestingly, our numerical 
studies give a negative value of $\theta$ in two and three dimensions.

The spiral trees are anisotropic. We measure the average extent of a
$n$-site spiral tree in $x-y$ plane and perpendicular to the $x-y$
plane through the moment of inertia tensors, $I_{pl,n}$ and
$I_{\perp,n}$ respectively. In the asymptotic limit, they are expected
to vary as

\begin{equation}
I_{pl,n} \sim A n^{2 \nu_{pl}+1}
\end{equation}
and

\begin{equation}
I_{\perp,n} \sim A n^{2\nu_{\perp}+1}
\end{equation}
where $\nu_{pl}$ and $\nu_{\perp}$ define the length scale of the
spiral tree in planar and perpendicular direction respectively.

\section{Two dimensional lattice spiral trees}

Some pictures of randomly generated large spiral trees are shown in
Fig. \ref{2dmc} (details later). One notes very long one dimensional
structures  with infrequent turns. Hence, simple counting of
structures of kind shown  in Fig.\ref{backb} should give a good
estimate of the growth constant  $\lambda$. The generating function of
trees of this type is easy to  determine. If $A_1(x)$ is the
generating function, we get

\begin{equation}
A_1(x) =  \frac{x}{1-x} + \frac{x^3}{(1-x)^2} A_1(x)
\end{equation}
which gives $A_1(x) = \frac{x(1-x)}{1+x^2-2x-x^3}$. The number of
trees of  this type grows as ${\lambda_1}^n$, with $\lambda_1
=1.754878$.  It is straightforward to include more complicated
branches in such counting  to get a better lower bound. This we
proceed to do below.

\subsection{Lower Bound on Growth Constant on Square lattice}

Consider a subset of all the spiral trees on a square lattice rooted
at the origin, which lie strictly in the first quadrant  $x \geq 0$,
$y \geq 0$; starting at the origin, and not touching $y=0$ and  $y=1$
except at points $(0,0)$ and $(0,1)$ respectively. If $Q(x)$ is the
generating function for spiral trees in a quadrant and if $q_{4,n}$ is
the coefficient of $x^n$ in the expansion of $([Q(x)]^4)/x^3$, then

\begin{equation}
A_n \ge q_{4,n}
\end{equation}
where $A_n$ is the $n^{th}$ term of $A(x)$, the generating function of
all  spiral trees on the square lattice.

The enumeration of graphs contributing to $Q(x)$ can be made easier
by noticing that these graphs can be formed by combination of smaller
graphs. We define an articulation point \cite{essam} as a point on
$y$-axis such that the tree above is an allowed spiral tree in the
quadrant above that part (note that these trees are defined in the 
upper quadrant and they never touch $y=0$ axis, except at (0,0)). For 
example, the solid squares represent the articulation points of the 
graph in Fig. \ref{backb}, and Fig.\ref{ni} shows a spiral tree with 
no articulation point. Hence, these spiral trees can be seen as trees 
having $y$ axis as a backbone on which spiral graphs are connected at 
different articulation points maintaining the spiral constraint.

Let $B(x)$ be the generating function of the quadrant spiral trees
with  no articulation points. Hence $B(x)$ can be seen as sum of
generating  function of irreducible graphs with $i$ sites along
$y$-axis. We represent  them by $B_i(x)$ (see Fig. \ref{bix}), then
$B(x) =  \sum_{i=1}^{\infty} B_i(x)$. The full generating function in
terms  of $B(x)$ would be

\begin{equation}
Q(x) = x(1+ B(x)+ B^2(x).......) = \frac{x}{1-B(x)}
\end{equation}
where $B_i(x)$ are spiral graphs starting with $i$-sites along the
$y$-axis. It is easy to see that $B_1(x)=x$, $B_2(x)= \frac{x^3}{1-x}$
and $B_3(x) = \frac{x^6}{(1-x-x^3)(1-x)}$. One can write $B_4(x)$ with
some effort but we do not have a general form for $B_i(x)$ for any
$i$.

We restrict the graphs contributing to $B_i(x)$ to be graphs such that
they have $i$ sites along the $y$ axis and have at least one downward
branch with $i-1$ sites. This would not include structures like Fig.
\ref{ni}. We will represent the generating function of these graphs
by $Q_1(x)$. Then we can represent $B_i(x)$ in terms of two other
generating functions, $V_i(x)$ and $W_i(x)$. We define $V_i(x)$ as the
generating function of spiral subgraphs starting with having $i$ sites
along $y$-axis. $W_i(x)$ is the generating function of spiral
subgraphs  staring with $i$-sites along $y$-axis and ending with a
downward branch  with $i-1$ sites (Fig.\ref{wv}). Then,

\begin{equation}
B_i(x) = W_i(x)+\frac{W_i(x) V_{i-1}(x)}{x^{i-1}}  \label{Bi}
\end{equation}

Also, $V_i(x)$ can be rewritten in terms of $W_i(x)$ as

\begin{equation}
V_i(x) = x V_{i-1}(x)+ W_i(x)+ \frac{W_i(x) V_{i-1}(x)}{x^{i-1}}
\end{equation}

By expressing $Q_1(x)$ in terms of $B_i(x)$ and $B_i(x)$ in turn in
terms of $W_i(x)$, we can reduce the computational time. If $W_{n}$
is the number of graphs of size $n$ contributing to $W(x)$ ($W(x) =
\sum_{i=1}^{\infty} W_i(x)$), and $Q_n$ is the number of graphs of
size  $n$ contributing to $Q_1(x)$, then  $W_n$ grows more slowly than
$Q_n$. We  enumerated $W_n$ and using them we could generate a $56$
term series for  $Q_1(x)$. The computation time for $W_n$ grows as
$(1.8)^n$, in contrast to  $(2.04)^n$ for the $Q_n$ series.

If we restrict the graphs contributing to $B_i(x)$, $W_i(x)$ and
$V_i(x)$ to  the graphs having comb-like structure (by comb-like 
structure we mean graphs with one dimensional backbone having 
vertical straight lines of arbitrary lengths), then it turns out
that one can  get the exact expression for these generating
functions. We represent them by
$\widetilde{V}_i(x),\widetilde{W}_i(x)$ and $\widetilde{B}_i(x)$. It
is easy to  see that for comb like structures,

\begin{equation}
W_i(x) \geq \widetilde{W}_i(x)= \frac{x^{2i}}{1-x} +\frac{x^{2i}}{1-x}
\frac{K(x)}{1-x}+\frac{x^{2i}}{1-x}
\left(\frac{K(x)}{1-x}\right)^2+.....
\end{equation}
where $K(x) =  x^2 \sum_{j=1}^{i-2} x^j$.  Hence,

\begin{equation}
\widetilde{W}_i(x) = \frac{x^{2 i} (1-x)}{1-2x+x^2-x^3+x^{i+1}}
\end{equation}

Similarly, we get

\begin{equation}
\widetilde{V}_i(x) = \frac{x^{i+1} (1-x+x^2-x^i)}{1-2x+x^2-x^3+x^{i+1}}
\end{equation}
and hence

\begin{equation}
B_i(x) \geq \widetilde{B}_i(x) = \frac{x^{2i}
(1-x)^2}{(1-2x+x^2-x^3+x^i)(1-2x+x^2-x^3+x^{i+1})}
\end{equation}

Substituting in Eq. \ref{Bi} we get the generating function,
$\widetilde{Q}_1(x)$ for this subset of spiral trees in a
quadrant. This  generating function has a singularity at $x_c =0.51662
$ which  gives the growth constant $\lambda^{'}$ of these trees to be
$1.93565$.  Since this counts only a subset of all the spiral trees on
a square lattice,  this is a rigorous lower bound on
$\lambda_{spiral}$ for spiral  trees on a square lattice.

For the full $Q_1(x)$, we derived a 56 term series. If we assume,

\begin{equation}
Q_{n} \sim \lambda_1^n n^{-\theta_1}
\end{equation}
then we got  estimates of $\lambda_1$ and $\theta_1$ to be

\begin{eqnarray} 
\lambda_1 = 2.0449 \pm 0.0001\\ \theta_1 =  0.830 \pm 0.01
\end{eqnarray}

\subsection{Exact enumeration}

Since the number of configurations of a given cluster size is
exponential in cluster size, the computational complexity of the
algorithm for enumeration of all lattice animals or trees grows
exponentially with the cluster size. For direct enumeration algorithms
like Martin's algorithm \cite{martin}, the time required to generate
all the configurations of a given size grows as $\lambda^n$, where
$\lambda$ is the growth constant and $n$ is the cluster size and the
memory requirement grows like a polynomial in cluster size. For
lattice trees and animals, a finite lattice method  \cite{enting} with
an associated transfer matrix algorithm was used  by Conway
\cite{conway}. Conway generated a $25$ term series for  lattice
animals using this algorithm. This series has recently  been extended
to $46$ terms by Jensen \cite{jensen} with a slight improvement in the
algorithm. Both space and time requirements of this algorithm are
found numerically to approximately  grow as $1.4^n$. The growth
constant of lattice animals in contrast is approximately $4.06$. Hence
a considerable improvement in time is obtained by the transfer matrix
algorithm at the cost of memory.

The spiral constraint, cannot be easily implemented using the transfer
matrix. Hence we have used Martin's algorithm for spiral trees,
making use of the four-fold rotational symmetry of the lattice. Our
series  for number of trees and their average moment of inertia is
given in  Appendix.1.

Using this we generated a series of spiral trees on square lattice
up to  37 terms (Appendix 1). Earlier known series had only $25$ terms.

For analysing the series data we tried a four parameter sequential fit
to the data of the form
 
\begin{equation}
A_n =  B {\lambda}^n (n+\delta)^{-\theta} \label{seqa}
\end{equation}
where $\delta$ is an adjustable fixed parameter and $B$ is a constant. 
We did a linear fit on the logarithm of Eq. \ref{seqa} using $A_n$, 
$A_{n+1}$, $A_{n+2}$ and $A_{n+3}$ to estimate values of
$B_n$, $\delta_n$, $\lambda_n$ and $\theta_n$. For spiral trees on
square lattice we found a good convergence in successive values of
$\lambda_n$ and $\theta_n$ for $\delta$ lying between 2.03 and 2.04.
Fixing $\delta =2.0367$ and $B=0.18124$ we get a very good convergence
of $\lambda_n$ and $\theta_n$ for different values of $n$. These are
given in Appendix 2. From this we estimate

\begin{eqnarray}
\lambda = 2.11433\pm0.00010\\ \theta = -1.3667 \pm 0.0010
\end{eqnarray}

We have tried fits with non analytic corrections to scaling of the form , 
$B {\lambda}^n (n+\delta)^{-\theta} [1+a/n^{\Delta}]$, but we didn't get good convergence for $\Delta$. Instead, $B {\lambda}^n (n+\delta)^{-\theta} [1-\alpha e^{-\beta n} ]$ seems to fit much better with $\alpha \approx 0.32$  and $\beta \approx 0.35$.

For the radius of gyration data we used a sequential fit of the form

\begin{equation}
\mbox{log}I_{i,n}  = (2\nu_i+1) \mbox{ln}(n+\delta)+ u
+\frac{v}{(n+\delta)^2} \label{seqr}
\end{equation} 
where $i$ stands for $pl$ or $\perp$ as the case maybe and $u$ and $v$
are constants.

For spiral trees in a plane $I_{\perp,n}$ would be zero and by
symmetry  the sum of square of $x$ coordinate of all sites for all
configurations  of clusters of size $n$ is symmetric with sum of
squares of $y$-coordinate.  Using Eq. \ref{seqr} for sequential fit to
our $35$ term series we get a  good convergence for $\delta$ lying
between $-0.33$ and $-0.35$. Fixing  $\delta =-0.338$ we get the
estimates of $\nu_{pl}$ to be

\begin{equation}
2 \nu_{pl} = 1.3148 \pm 0.0010\\
\end{equation}  

These estimates are much more precise than the earlier estimates
$\lambda =2.1166 \pm 0.001$, $\theta = -1.307 \pm 0.006$ and  $2
\nu_{pl} = 1.306 \pm 0.010$ using a $25$ term series \cite{bose}. We 
can rule out the dimensional reduction conjecture with fair confidence.

Above we presented our estimates using four parameter fits. Method of
differential approximants has almost become a standard technique for
such analysis \cite{guttmannp}. In this case, the generating function 
has a divergent singularity at $x_c$. We tried zeroth order differential 
approximants, they are listed in Table 1. We find a very poor convergence 
in values of $x_c$ and $\theta$. Out of $70$ approximants, $15$ show 
spurious singularities (singularities with $|x_c| < 0.45)$. We have 
listed $20$ values which showed best convergence. From these we get, 
$\lambda = 2.1142 \pm 0.002$ and $\theta = -1.39
\pm 0.02$. Clearly the series is not very well behaved. This is
reflected in the slow convergence of our series. Also Monte-Carlo
generated random spiral trees of sizes $1000$ (Fig. \ref{2dmc})
suggest that the asymptotic behaviour of the series might set in
rather late. Because of poor convergence of differential approximants,
we have relied on parameter fits for series analysis in this paper.

\subsection{Monte-Carlo analysis}

\begin{figure}
\begin{center}
		 \epsfig{figure=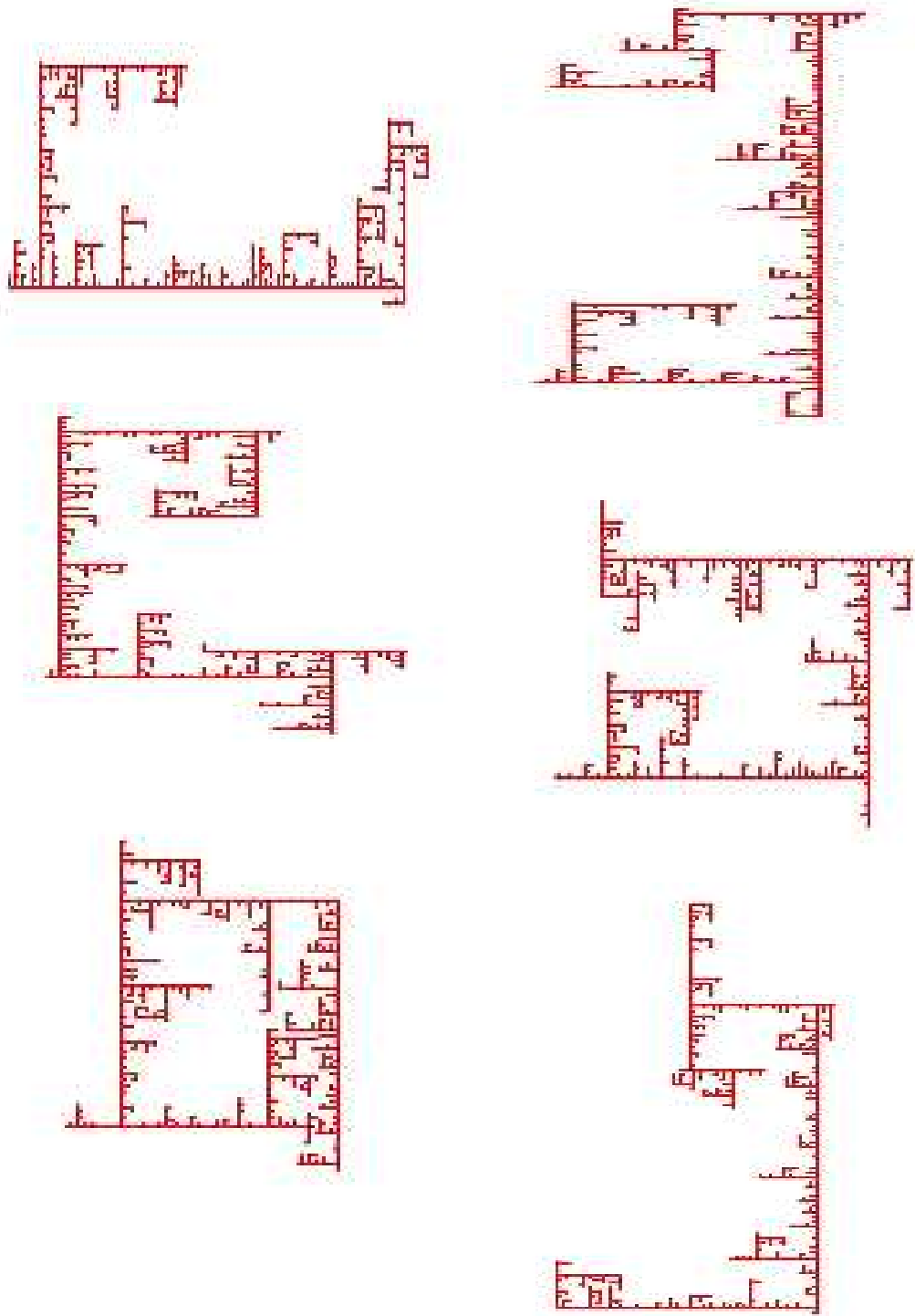,height=20cm}
\end{center}
\caption{Randomly generated spiral trees of 1000 sites in 2-dimensions
using incomplete-enumeration algorithm}  \label{2dmc}
\end{figure}

\begin{figure}
\begin{center}
		 \epsfig{figure=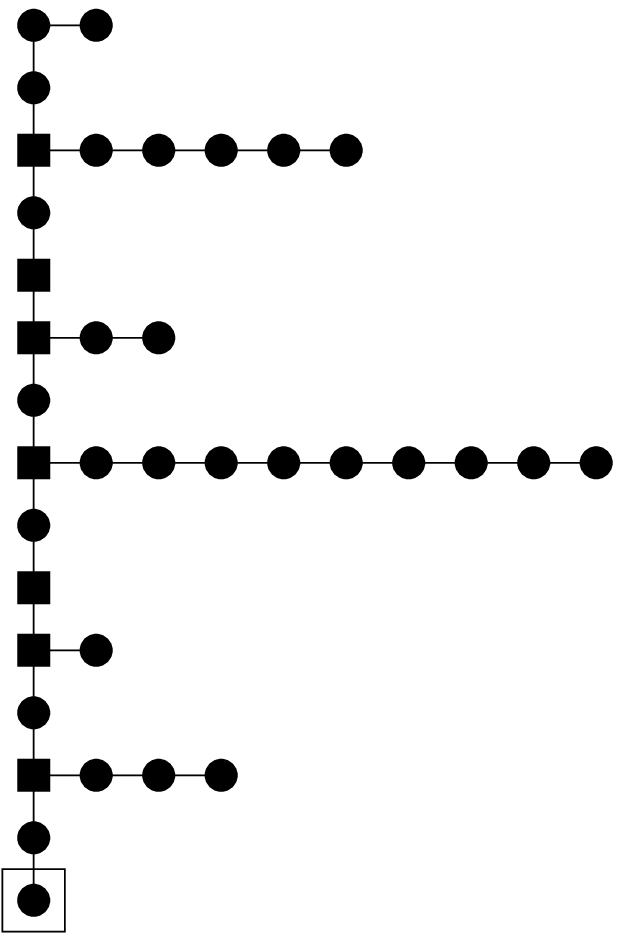,height=6cm}
\end{center}
\caption{A simple counting problem of backbone with arbitrary long
offshoots. Minimum distance between two offshoots is $2$ as else  the
tree constraint is violated. Solid squares represent the  articulation
points of the graph.}   \label{backb}
\end{figure}

\begin{figure}
\begin{center}
		 \epsfig{figure=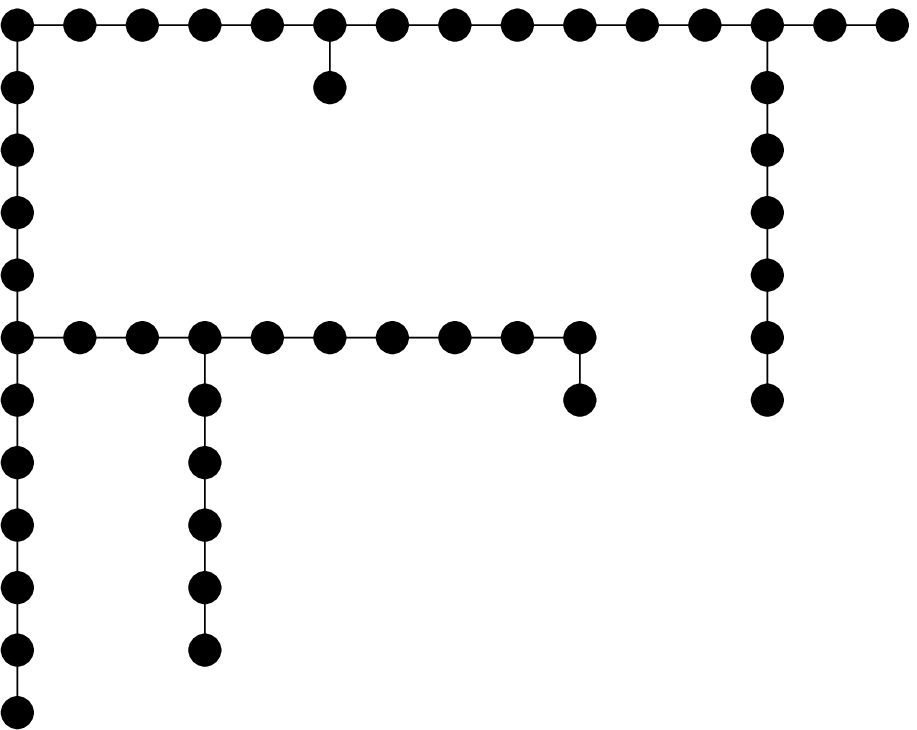,height=7cm}
\end{center}
\caption{An example of an irreducible spiral graph with no
articulation  point. This is also an example of a graph not included
in $Q_1(x)$} \label{ni}
\end{figure}

\begin{figure}
\begin{center}
\epsfig{figure=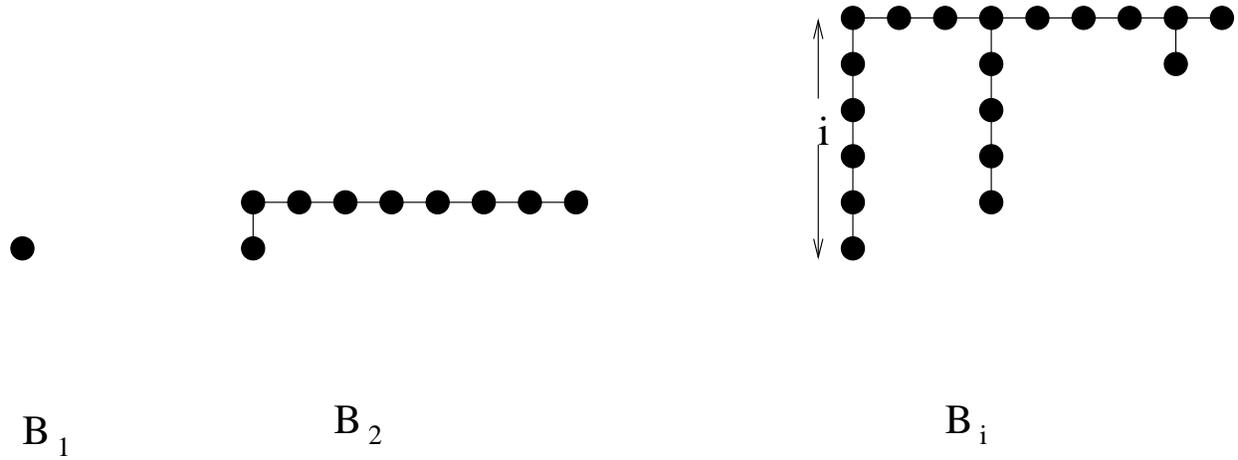,height=6cm}
\end{center}
\caption{Schematic figure of spiral trees contributing to $B_i(x)$.
$B_1(x)$ is just a  single vertex.} \label{bix}
\end{figure}

\begin{figure}
\begin{center}
\epsfig{figure=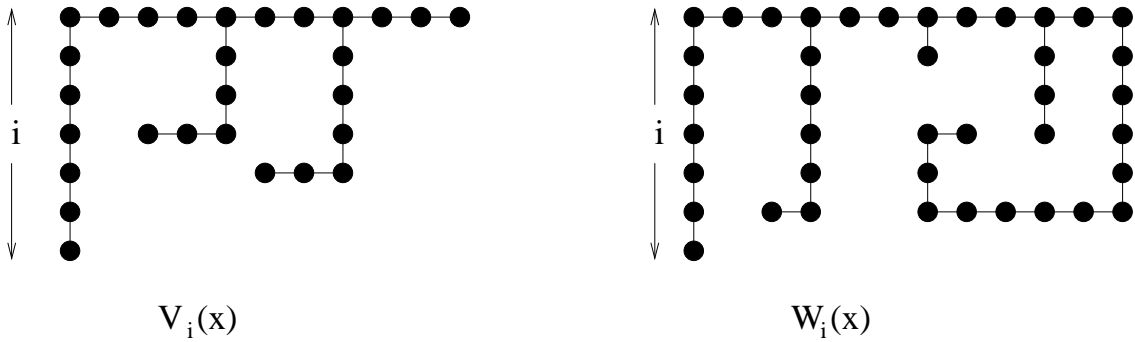,width=15cm}
\end{center}
\caption{Example of graphs contributing to $V_i(x)$ and $W_i(x)$
respectively.} \label{wv}
\end{figure}

With exact enumeration, we are restricted to clusters of size thirty
seven in two dimensions. The main problem is with the extrapolation
since the crossover sizes are likely to be large, as the total angle 
turned by the largest spiral arm about the origin for a spiral tree 
of size $1000$ is about $2 \pi$ only (Fig. \ref{2dmc}). This indicates 
that the crossover value above which asymptotic behaviour sets in would 
be of order $10^3$. We tried to study larger spiral trees using MC 
methods. Monte-Carlo simulation of branched polymers is a challenging 
problem. Because of branching, most MC algorithms which are good for 
linear polymers show critical slowing down for branched polymers. For 
lattice trees there have been some studies using the cut and paste 
dynamic MC technique \cite{madras}. But with spiral constraint, 
algorithms involving large scale non local moves are not useful. We used 
an improved version of incomplete enumeration algorithm proposed recently by us
\cite{sumedha}. Using it we could study spiral trees of  sizes up to
one thousand on a square lattice.

Incomplete enumeration is a simple modification of exact enumeration
algorithm and can be seen as a percolation process on the genealogical
tree of the underlying enumeration problem. The optimal behaviour of
the algorithm is achieved when we work around the percolation threshold
of the genealogical tree. This algorithm falls in the class of
stochastic growth algorithm like PERM \cite{grassberger}. We have
shown in \cite{sumedha}, that the asymptotic time to produce an independent
sample of $n$ sites  for trees and animals grows as $exp(a n^b)$ with 
$0 <b<1$ for this algorithm. Though the coefficient in front of stretched 
exponential can be made small by optimising the algorithm for small
sizes. We will not give more details of the algorithm in this paper. 
These can be found in \cite{sumedha}.

Fig.\ref{2dmc} shows pictures of some typical spiral trees of one
thousand sites. Clearly, their structure is very different from
lattice  trees without the spiral constraint. Because of the constraint
they tend to  branch much less. For spiral constraint, earlier
numerical evidence  suggest that unlike lattice trees and animals,
spiral trees and animals  do not lie in same universality class. The
reason is that by allowing  loops, the polymer can bend much more
often and hence spiral animals  would be more compact than the spiral
trees.

\begin{figure}
\begin{center}
		 \epsfig{figure=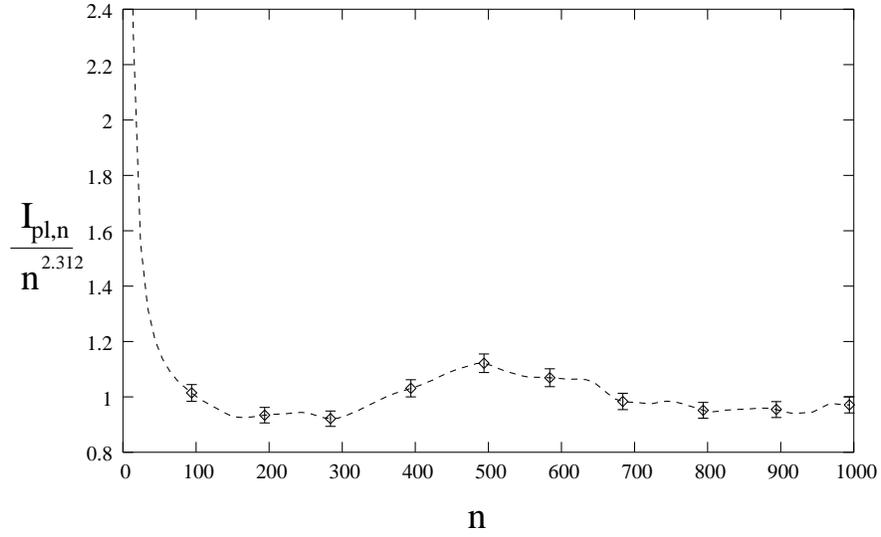,height=7cm}
\end{center}
\caption{Plot of $\frac{I_{pl,n}}{n^{2.312}}$ as a function of $n$ for
Monte-Carlo generated spiral trees on a square lattice.}
\label{2drad1}
\end{figure}

\begin{figure}
\begin{center}
		 \epsfig{figure=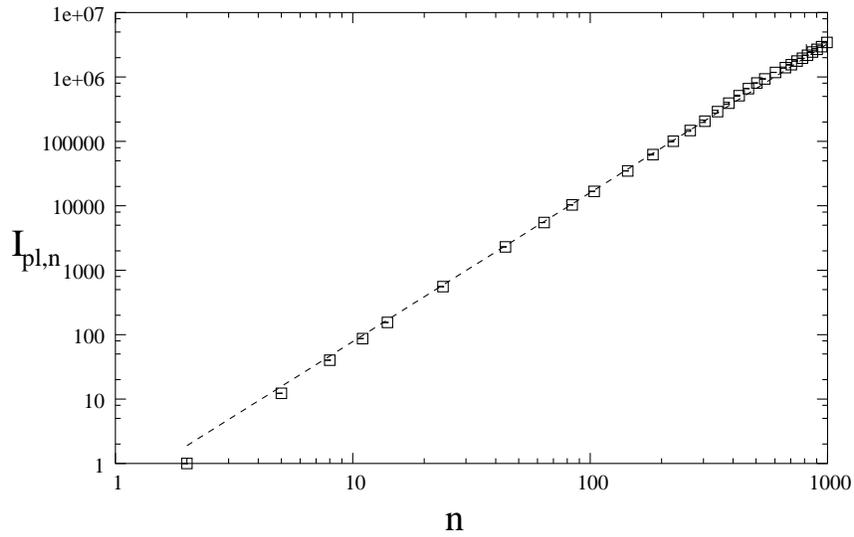,height=7cm}
\end{center}
\caption{Plot of $I_{pl,n}$ Vs $n$ for Monte-Carlo generated spiral
trees  on a square lattice. The dotted line is a straight line with
slope $2.312$.}
\label{2drad2}
\end{figure}

We studied spiral trees up to sizes $1000$ using incomplete enumeration
MC method. We made $10^7$ MC runs. The moment of inertia tensor
$I_{pl,n}$ as a function of $n$ is plotted in Fig \ref{2drad1} and Fig
\ref{2drad2}. Assuming the asymptotic form to be such that

\begin{equation}
\mbox{log}(I_{pl,n}) = \mbox{log}C +(2
\nu_{pl}+1)\mbox{log}n+\frac{D}{n}
\end{equation}

Using above written form, we get the estimate of $\nu_{pl}$ to be (Fig
\ref{2drad1} and Fig \ref{2drad2})

\begin{equation}
2\nu_{pl} =  1.312\pm 0.010
\end{equation}

\begin{figure}
\begin{center}
		 \epsfig{figure=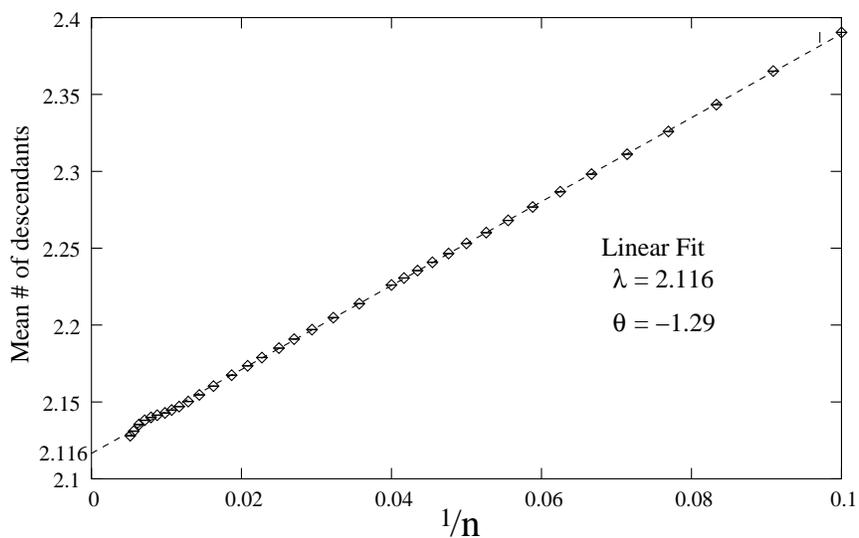,height=7cm}
\end{center}
\caption{Monte-Carlo estimates of ratios of the number of
configurations  on a square lattice. The straight line gives a linear
fit of the form $\lambda(1-\theta/n)$ to the data.} \label{ratio}
\end{figure}

In incomplete enumeration MC algorithm \cite{sumedha}, each
configuration  of $n$ sites is generated with equal probability $P_n$
which is just  $\prod_{i=1}^{n} p_i$, where $p_i$ is the probability
with which an edge  between level $i$ and $i+1$ on the genealogical
tree of the problem is  chosen. By keeping track of the average number
of clusters of a given  size generated in a given run, one can
estimate the growth constant  $\lambda$ and the critical exponent
$\theta$. But, the variance of the  number of clusters increases as
$\mbox{exp}(n^{\alpha})$, $0<\alpha<1$ for  large $n$. Hence, instead
we counted the number of descendants of each  spiral tree
generated. This approach has been used previously in  \cite{bose,rensburg}. 
The mean number of descendants of a  tree of size $n$ gives a direct 
estimate of $A_{n+1}/A_n$. We represent the  mean number of descendants 
by $M_n$. This is plotted in Fig. \ref{ratio}. A linear fit of the  
form $\lambda (1-\theta/n)$ to this data gives $\lambda =2.116 \pm 0.01$ 
and $\theta = -1.29\pm 0.02$. For better estimates we assume

\begin{equation}
\mbox{log}M_n = \mbox{log}\lambda-\theta
\mbox{log}\left(\frac{n+\delta}{n-1+\delta}\right)
\end{equation}

With this we get the following estimates for $n\le 200$ which are in 
agreement with the value  obtained by extrapolating the exact series 
expansions.
\begin{eqnarray}
\lambda = 2.1145 \pm 0.0010\\ \theta  =  -1.364 \pm 0.010
\end{eqnarray}
with $\delta=1.8$.

\section{Spiral trees on a cubic lattice}

In dimensions higher than two, the spiral constraint defined as the
projection of path joining any site of the tree to the root in $x-y$
plane contains no left turn can be employed in two ways. Bose
et. al. \cite{bose} defined it such that for the projected path from
origin to site on $x-y$ plane only forward and right turns are
allowed.  But in dimensions higher than two, we can define another
variation where  trees as long as they do not violate the tree
constraint and the projection  on $x-y$ plane is spiral, are
allowed. We will call the spiral trees with only  forward and right
turns allowed as $ST_1$.

\begin{figure}
\begin{center}
		 \epsfig{figure=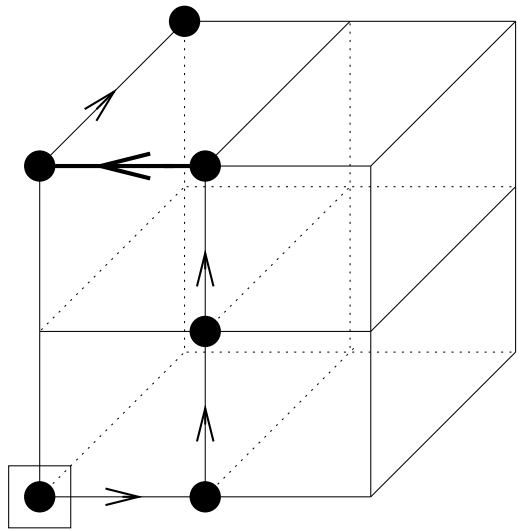,height=7cm}
\end{center}
\caption{A spiral tree of six sites on a cubic lattice with a
back-turn (drawn by a thicker line). This configuration will
contribute to spiral trees $ST_2$ of six sites but not to $ST_1$.}
\label{3dst}
\end{figure}

If we allow for back-turns also, we would get different series because
for  example, Fig \ref{3dst} shows one spiral tree of six sites which
would not  be a valid configuration if we consider only forward and
right turns.  We call the spiral trees with back-turns allowed as
$ST_2$. Naively, one  would expect these two to belong to the same
universality  class. We generated the series till $n=17$ on a cubic
lattice using both  definitions, however we find the two series
behaving differently.  Series for both $ST_1$ and $ST_2$ are given in
Appendix 1.

For $ST_1$, for $A_n$ the number of configurations, using Eq.
\ref{seqa} we find that the sequential fit shows a good convergence
around $\delta =2.43$. With $\delta =2.43$ and $B = 0.094$, the values
of  $\lambda$ and $\theta$ obtained are listed in Table 1. For
$\nu_{pl}$ and  $\nu_{\perp}$, we used fitting form as given in
Eq. \ref{seqr}, with  $\delta =-1.46$ and $\delta =-0.43$
respectively. The sequential fits are  given in Appendix 2 and
estimates are listed in Table 2.

Similarly, we obtained $17$ term series for $ST_2$. The sequential
fits  are given in Appendix 2 and the values of $\lambda$, $\theta$,
$\nu_{pl}$  and $\nu_{\perp}$ are listed in Table 2.

The difference in value of $\lambda$ for $ST_1$ and $ST_2$ is
understandable as $ST_2$ has a greater number of configurations. More
surprisingly, the critical exponents $\theta$, $\nu_{pl}$ and
$\nu_{\perp}$ within our error estimates are different in two models.
In neither case, the conjectured dimensional reduction(Eq.1 and 2)
seems  to be satisfied.

\section{Spiral trees in four dimensions}

On a hyper cubic lattice in four dimensions we generated a series till
$n=13$. We also correct mistakes in the earlier series reported for
$ST_1$ in \cite{bose}. The corrected series in given in the Appendix
1.  We also obtained a 13 term series for $ST_2$(see Appendix 1). The
estimates  of $\lambda$ and critical exponents are listed in Table 2.

We also performed Monte-Carlo simulations using incomplete enumeration
algorithm for spiral trees up to size $50$. Our estimates from MC
simulations for $ST_1$ are given in Table 3.

Though we cannot rule out the possibility of $\theta$ being zero in
both  series analysis and Monte-Carlo simulations, but it seems
unlikely.

\section{Discussion}
Bose et. al. gave a plausible argument of curling up of the dimensions 
in the spiralling plane and had conjectured a dimensional reduction 
by four for spiral trees. Our numerical evidence as presented in 
this paper does not support the conjecture. The
spiral  constraint for trees seems to be very special. For example,
the  structure of spiral trees is very different from spiral animals
with  loops allowed \cite{bose1}. Different implementation of the 
constraint in $d>2$,  seems to give different critical behaviour, suggesting
different universality classes. A variety of self avoiding walks with
different step restrictions rules on simple cubic lattice were studied
in \cite{asaw} using exact enumeration. Their analysis suggested same 
universality class for self avoiding walks with various restrictions 
(including the spiral constraint), as the unrestricted self avoiding walks. 
In contrast, our studies show different critical behaviour of spiral trees 
with different geometrical restrictions in three and four dimensions.

We should note that for the large clusters of size $10^3$ generated
by Monte Carlo, the total angle turned by the largest spiral arm about
the origin is about $2 \pi$. It is possible that the structure of
spiral trees is such that this angle tends to infinity as $n$ tends to
infinity. In this case the crossover value above which asymptotic
behaviour sets in would be expected to be of order $10^3$, and series
analysis for smaller $n$ may not give correct limiting behaviour. One
indication that trees where spiral turns a lot are important is that
the growth constant for spiral trees in a quadrant $Q_1(x)$ seems to
be significantly smaller than for full spiral trees.

For quadrant spiral trees on a square lattice, we obtained exact
series  up to sizes 56. There are very few such long series known for
lattice models.  The series gives a estimate of $\lambda=2.044$ for
these quadrant spiral  trees. This value is significantly smaller than
for the full spiral trees.

\section{Acknowledgements}
I am thankful to my adviser Prof. Deepak Dhar for suggesting the
problem,  for discussions and for useful comments on the manuscript.

\section{Appendix 1}
Exact series enumeration values in different dimensions

\subsection{Two dimensions}

\begin{tabular}{|l|l|l|} \hline
cluster size(n) & $A_n$ & $\langle{I_{pl,n}}\rangle$ \\ \hline 1   &
1         &         0\\ \hline 2   &    4         &         1\\ \hline
3   &    14        &         3.142857\\ \hline 4   &    40        &
6.800000\\ \hline 5   &    105       &         12.266667\\ \hline 6
&    268       &         19.656716\\ \hline 7   &    674       &
28.919881\\ \hline 8   &    1660      &        40.159036\\ \hline 9
&    4021      &         53.513056\\ \hline 10  &    9612      &
69.074906\\ \hline 11  &    22734     &         86.926014\\ \hline 12
&    53276     &         107.140851\\ \hline 13  &    123916    &
129.787372\\ \hline 14   &    286376    &         154.926432\\ \hline
15   &    658100    &         182.624835\\ \hline 16   &    1504900
&        212.938547\\ \hline 17   &   3426464     &       245.919131\\
\hline 18   &   7771444      &      281.619675\\ \hline 19   &
17565064     &      320.089299\\ \hline 20   &   39576360      &
361.374917\\ \hline 21   &   88916877      &     405.522760\\ \hline
22   &   199252252     &     452.577078\\ \hline 23   &   445438310
&     502.580546\\ \hline 24   &   993616344      &    555.575100\\
\hline 25   &   2211923712     &    611.601183\\ \hline 26   &
4914811468    &     670.697934\\ \hline 27   &  10901498938    &
732.903853\\ \hline 28   &  24141259980    &    798.256392\\ \hline 29
&   53379537257   &     866.791847\\ \hline 30   &  117861710196    &
938.545859\\ \hline 31   &   259891311248   &    1013.553288\\ \hline
32   &   572356464452   &    1091.848086\\ \hline 33   &
1259008971656  &    1173.463504\\ \hline 34   &   2766351037428  &
1258.432171\\ \hline 35   &   6071954146120  &    1346.786006\\ \hline
36   &  13314252070412  &   \\ \hline 37   &  29167189621351  &   \\
\hline
\end{tabular}

\subsection{Three Dimensions $ST_1$}

\begin{tabular}{|l|l|l|l|} \hline
$n$ & $A_n$ & $\langle{I_{pl,n}}\rangle$ &
$\langle{I_{\perp,n}}\rangle$ \\ \hline 1 & 1 & 0.& 0.\\ \hline 2 & 6
&  0.66666 & 0.333333\\ \hline 3& 41             &  1.85366 &
1.07317\\ \hline 4& 260            &  3.63076 & 2.27692\\ \hline 5&
1568           &  6.02296 & 3.98214\\ \hline 6& 9190           &
9.06464 & 6.19913\\ \hline 7& 53090          & 12.75954 & 8.91987\\
\hline 8& 303900         & 17.09588 & 12.1405\\ \hline 9& 1727691
& 22.0606 & 15.8606\\ \hline 10& 9767426       & 27.6424 & 20.0821\\
\hline 11& 54966550      & 33.8322 & 24.8071\\ \hline 12& 308138528
& 40.6214 & 30.0376\\ \hline 13& 1721739000    & 48.0022 & 35.7754\\
\hline 14& 9592901762    & 55.9676 & 42.0229\\ \hline 15& 53314247488
& 64.5112 & 48.7822\\ \hline 16& 295644339728  & 73.6274 & 56.0556\\
\hline 17& 1636179620652 & 83.3112 & 63.8454\\ \hline
\end{tabular}

\subsection{Three Dimensions $ST_2$}

\begin{tabular}{|l|l|l|l|} \hline
$n$ & $A_n$ & $\langle{I_{pl,n}}\rangle$ &
$\langle{I_{\perp,n}}\rangle$ \\ \hline 1 & 1              & 0 & 0.\\
\hline 2 & 6              & 0.666666 & 0.333333\\ \hline 3 & 41 &
1.85366 & 1.07317\\ \hline 4 & 260            & 3.63076 & 2.27692\\
\hline 5 & 1576           & 6.00762 & 4.00761\\ \hline 6 & 9342 &
9.00192 & 6.30208\\ \hline 7 & 54890          & 12.60084 & 9.17041\\
\hline 8 & 320952         & 16.7848  & 12.6182\\ \hline 9 & 1869907 &
21.5398 & 16.651\\ \hline 10 & 10861750      & 26.8572 & 21.2772\\
\hline 11 & 62939998      & 32.7312 & 26.5047\\ \hline 12 & 364004296
& 39.156 & 32.3409\\ \hline 13 & 2101795408    & 46.1276 & 38.7927\\
\hline 14 & 12119643810   & 53.6422 & 45.8667\\ \hline 15 &
69805863940   & 61.6968 & 53.5693\\ \hline 16 & 401668665200  &
70.2898 & 61.9068\\ \hline 17 & 2309283532000 & 79.4192 & 70.8851\\
\hline
\end{tabular}

\subsection{Four Dimensions $ST_1$}

\begin{tabular}{|l|l|l|l|} \hline
$n$ & $A_n$ & $\langle{I_{pl,n}}\rangle$ &
$\langle{I_{\perp,n}}\rangle$ \\ \hline 1 & 1               & 0. &
0\\ \hline 2 &  8              & 0.5   &0.5\\ \hline 3  & 80
&  1.35  & 1.5\\ \hline 4  & 800            &  2.54  & 3.030\\ \hline
5   &7912           &  4.05864 &  5.10010\\ \hline 6   &77656
&  5.89816 &   7.70862\\ \hline 7   &759172         & 8.04822   &
10.84584\\ \hline 8 &  7403292        & 10.49742  &  14.50268\\ \hline
9  & 72073417       & 13.23410  &  18.67008\\ \hline 10  & 700774524
& 16.24692  &  23.34\\ \hline 11  & 6806914432    & 19.52526  &
28.5052\\ \hline 12  & 66064406668   & 23.0592   &  34.1596\\ \hline
13  & 640741734643  & 26.8396   &  40.2974\\ \hline
\end{tabular}

\subsection{Four dimensions $ST_2$}

\begin{tabular}{|l|l|l|l|} \hline
$n$ & $A_n$ & $\langle{I_{pl,n}}\rangle$ &
$\langle{I_{\perp,n}}\rangle$ \\ \hline 1 &  1             &  0.  &
0.\\ \hline 2  & 8             &  0.5  & 0.5\\ \hline 3 &  80
&  1.35  & 1.5\\ \hline 4  & 800           & 2.54  & 3.030\\ \hline 5
& 7960          & 4.05226  & 5.10754\\ \hline 6  & 79048         &
5.87628  &  7.74208\\ \hline 7  & 785748        & 7.99822 &
10.93174\\ \hline 8  & 7822676       & 10.40506 &  14.6724\\ \hline 9
& 78011513      & 13.08484  & 18.95778\\ \hline 10  & 779189988    &
16.0274  & 23.7816\\ \hline 11  & 7793589943   & 19.22410 &  29.1376\\
\hline 12  & 78049537766  & 22.6676  & 35.0206\\ \hline 13 &
782489000000 &  26.3518  & 41.4252\\ \hline
\end{tabular}

\section{Appendix 2}
\subsection{Two Dimensions Sequential Fit}

\begin{tabular}{|l|l|l|l|} \hline
 n & $\lambda_n$ & $\theta_n$ & $2 \nu_{pl,n}$ \\ \hline 5    &
 2.078982187 &-1.4143616  & 1.2918751\\ \hline 6  &    2.118727624
 &-1.3598402 & 1.3047319\\ \hline 7    &    2.117039314 & -1.3623751 &
 1.3092198\\ \hline 8  &    2.115352878 & -1.3651395 & 1.3108492\\
 \hline 9   &     2.114617151 & -1.3664433 & 1.3117861\\ \hline 10  &
 2.114771869 & -1.3661493 & 1.312420\\ \hline 11  &     2.113813740 &
 -1.3680905 & 1.3128895\\ \hline 12  &     2.113978775 & -1.3677359 &
 1.3132536\\ \hline 13  &     2.114183882 & -1.3672706 & 1.3135423\\
 \hline 14   &     2.114099443 & -1.3674721 & 1.3137672\\ \hline 15  &
 2.114103267 & -1.3674625 & 1.3139586\\ \hline 16   &     2.114205656
 & -1.3671946 & 1.3141194\\ \hline 17   &    2.114223238 & -1.3671466
 & 1.3142505\\ \hline 18   &     2.114256310 & -1.3670527 &
 1.3143596\\ \hline 19  &     2.114279786 & -1.3669834 & 1.3144497\\
 \hline 20  &     2.114291286 & -1.3669483 & 1.3145234\\ \hline 21  &
 2.114301033 & -1.3669174 & 1.3145839\\ \hline 22  &     2.114310834 &
 -1.3668854 & 1.3146334\\ \hline 23 &     2.114311487 & -1.3668832 &
 1.3146734\\ \hline 24 &     2.114314464 & -1.3668728 & 1.3147059\\
 \hline 25  &     2.114318963 & -1.3668566 & 1.3147321\\ \hline 26 &
 2.114320428 & -1.3668513 & 1.3147529\\ \hline 27 &     2.114321722 &
 -1.3668464 & 1.3147694\\ \hline 28 &     2.114324605 & -1.3668351 &
 1.3147823\\ \hline 29 &    2.114326551 & -1.3668274 & 1.3147921\\
 \hline 30   &    2.114327932 & -1.3668217 & 1.3147994\\ \hline 31
 &    2.114329734 & -1.3668142 & 1.3148047\\ \hline 32  &
 2.114331349 & -1.3668072 & 1.3148083\\ \hline 33    &    2.114332328
 & -1.3668029 & 1.3148104\\ \hline 34  &   2.114333055 & -1.3667997 &
 1.3148113\\ \hline 35     &  2.114333550 & -1.3667974 & 1.3148113\\
 \hline 36   &   2.114333553 & -1.3667974 & \\ \hline Es. Val. &
 $2.11433 \pm 0.0001$ & $-1.3667 \pm 0.001$ &  $1.3148 \pm 0.001$ \\
 \hline
\end{tabular}

\subsection{Three Dimensions $ST_1$}

\begin{tabular}{|l|l|l|l|l|} \hline
 n & $\lambda_n$ & $\theta_n$ & $2 \nu_{pl,n}$ & $2 \nu_{\perp,n}$\\
 \hline 5 & 5.153269 & -1.019107 & 0.847865 & 1.128949\\ \hline 6 &
 5.275382 & -0.810187 & 0.865641 & 1.098419\\ \hline 7 & 5.310873 &
 -0.743662 & 0.871330 & 1.083814\\ \hline 8 & 5.319667 & -0.725590 &
 0.874191 & 1.077873\\ \hline 9 & 5.327658 & -0.707695 & 0.875922 &
 1.073809\\ \hline 10 & 5.334141 & -0.691977 & 0.876525 & 1.070550\\
 \hline 11 & 5.337903 & -0.682161 & 0.876502 & 1.068326\\ \hline 12 &
 5.339533 & -0.677605 & 0.876303 & 1.067085\\ \hline 13 & 5.340111 &
 -0.675880 & 0.876176 & 1.066526\\ \hline 14 & 5.340282 & -0.675339 &
 0.876139 & 1.066334\\ \hline 15 & 5.340255 & -0.675428 & 0.876197 &
 1.066365\\ \hline Es. Val. & $5.340 \pm 0.02$ & $-0.675 \pm 0.05$ &
 $0.876 \pm 0.05$  & $1.066 \pm 0.05$\\ \hline

\end{tabular}

\subsection{Three Dimensions $ST_2$}

\begin{tabular}{|l|l|l|l|l|} \hline
 n & $\lambda_n$ & $\theta_n$ & $2 \nu_{pl,n}$ & $2 \nu_{\perp,n}$\\
 \hline 5  & 5.694072 & -0.143802 & 1.054721 & 1.200021\\ \hline 6  &
 5.719085 & -0.123718 & 1.013833 & 1.171160\\ \hline 7  & 5.710159 &
 -0.132441 & 0.989310 & 1.153823\\ \hline 8  & 5.695471 & -0.149408 &
 0.977105 & 1.147515\\ \hline 9  & 5.689143 & -0.157845 & 0.969817 &
 1.144294\\ \hline 10  & 5.687350 & -0.160552 & 0.963977 & 1.141637\\
 \hline 11  & 5.686110 & -0.162645 & 0.959561 & 1.139686\\ \hline 12
 & 5.684763 & -0.165153 & 0.956738 & 1.138646\\ \hline 13  & 5.683809
 & -0.167099 & 0.955250 & 1.138243\\ \hline 14  & 5.683473 & -0.167843
 & 0.954653 & 1.138124\\ \hline 15  & 5.683632 & -0.167463 & 0.954662
 & 1.138134\\ \hline Es. Val. & $5.683\pm0.02$ & $-0.167\pm0.05$ &
 $0.954\pm 0.05$  & $1.138\pm 0.05$\\ \hline

\end{tabular}

\pagebreak

\section*{Table Captions}

\begin{enumerate}

\item Table1:Estimates of critical exponents and growth constant from
differential approximants. We looked at approximants for $l \ge 9$ and 
$l-3 \le m \le l+3$. We have tabulated here $20$ values which showed best 
convergence.

\item Table2:Estimates of critical exponents and growth constant from
series  analysis in three and four dimensions. Note that the value of 
$\theta$ and $\nu$ for rooted lattice animals/trees in $3d$ and $4d$ is 
known exactly(in $3d$, $\theta=\nu=1/2$ and in $4d$, $\theta=5/6$ and 
$\nu=5/12$

\item Table3:Estimates of critical exponents and growth constants from
Monte-Carlo simulations in four dimensions.

\end{enumerate}

\pagebreak

\section*{Figure Captions}

\begin{enumerate}

\item Figure1:A rooted spiral tree of 15 sites on a  square lattice. The
root is the site enclosed in the square. At the root site the tree has
freedom of choosing any of the four neighbouring sites. We count the
spiral tree by number of sites and hence all bonds between two
occupied sites is always assumed to be present. The site marked by
$X$,  if present will result in a loop for spiral site trees and hence
will not  be allowed. But it can be present in a  spiral bond tree.

\item Figure2:Randomly generated spiral trees of 1000 sites in 2-dimensions
using incomplete-enumeration algorithm.

\item Figure3:A simple counting problem of backbone with arbitrary long
offshoots. Minimum distance between two offshoots is $2$ as else  the
tree constraint is violated. Solid squares represent the  articulation
points of the graph.

\item Figure4:An example of an irreducible spiral graph with no articulation 
point. This is also an example of a graph not included in $Q_1(x)$

\item Figure5:Schematic figure of spiral trees contributing to $B_i(x)$.
$B_1(x)$ is just a  single vertex.

\item Figure6:Example of graphs contributing to $V_i(x)$ and $W_i(x)$
respectively.

\item Figure7:Plot of $\frac{I_{pl,n}}{n^{2.312}}$ as a function of $n$ for
Monte-Carlo generated spiral trees on a square lattice.

\item Figure8:Plot of $I_{pl,n}$ Vs $n$ for Monte-Carlo generated spiral
trees  on a square lattice. The dotted line is a straight line with
slope $2.312$.

\item Figure9:Monte-Carlo estimates of ratios of the number of
configurations  on a square lattice. The straight line gives a linear
fit of the form $\lambda(1-\theta/n)$ to the data.

\item Figure10:A spiral tree of six sites on a cubic lattice with a
back-turn (drawn by a thicker line). This configuration will
contribute to spiral trees $ST_2$ of six sites but not to $ST_1$.

\end{enumerate}

\pagebreak

\begin{table}
\begin{center}
\begin{tabular}{|l|l|l|l|l|l|} \hline

$[l,m]$ &  $x_c = 1/\lambda$     & $\theta$  & $[l,m]$ & $x_c = 1/\lambda$ & $\theta$ \\ \hline 
$[14,13]$ & $0.47288256$ & $-1.36083$ & $[15,18]$ & $0.47307144$ & $-1.39078$   \\ \hline 
$[14,14]$ & $0.47290325$ & $-1.36384$ & $[14,17]$ & $0.47307308$ & $-1.39106$  \\ \hline 
$[15,13]$ & $0.47290516$ & $-1.36413$ & $[16,17]$ & $0.47307863$ & $-1.39209$  \\ \hline 
$[16,15]$ & $0.47294898$ & $-1.37035$ & $[17,15]$ & $0.47308675$ & $-1.39369$   \\ \hline 
$[13,15]$ & $0.47297513$ & $-1.37499$ & $[16,19]$ & $0.47309052$ & $-1.39421$    \\ \hline 
$[16,13]$ & $0.47303007$ & $-1.38409$ & $[18,15]$ & $0.47310355$ & $-1.39686$    \\ \hline 
$[13,16]$ & $0.47303305$ & $-1.38436$ & $[17,18]$ & $0.47310906$ & $-1.39788$    \\ \hline 
$[16,16]$ & $0.47305593$ & $-1.38800$ & $[15,16]$ & $0.47311001$ & $-1.39775$   \\ \hline 
$[14,15]$ & $0.47305793$ & $-1.38863$ & $[18,18]$ & $0.47311071$ & $-1.39822$  \\ \hline 
$[15,17]$ & $0.47306712$ & $-1.39002$ & $[17,19]$ & $0.47311091$ & $-1.39826$   \\ \hline 
\end{tabular}
\vspace{0.5cm}
\caption{Estimates of critical exponents and growth constant from
differential approximants. We looked at approximants for $l \ge 9$ and $l-3 \le m \le l+3$. We have tabulated here $20$ values which showed best convergence.}
\end{center}
\end{table}

\begin{table}
\begin{center}
\begin{tabular}{|l|l|l|l|l|} \hline

              & $ST_1 (d=3)$     & $ST_2 (d=3)$  & $ST_1 (d=4)$ &
$ST_2 (d=4)$ \\ \hline 
$\lambda$     &  $5.340 \pm 0.020$ & $5.683\pm 0.020$ & $9.62\pm 0.10$ & $10.20\pm0.10$ \\ \hline 
$\theta$ & $-0.675\pm 0.050$ & $-0.167\pm 0.050$ & $-0.11\pm 0.10$ & $0.29\pm 0.10$ \\ \hline 
$\nu_{pl}$    &  $0.44 \pm 0.05 $ & $0.477\pm 0.05$ &
$0.34\pm0.05$ &$0.37\pm 0.05$ \\ \hline $\nu_{\perp}$ &  $0.54 \pm
0.05 $ & $0.69\pm0.05$ &   $0.44\pm 0.05$ & $0.45\pm 0.05$\\ \hline
\end{tabular}
\vspace{0.5cm}
\caption{Estimates of critical exponents and growth constant from
series  analysis in three and four dimensions. Note that the value of $\theta$ and $\nu$ for rooted lattice animals/trees in $3d$ and $4d$ is known exactly(in $3d$, $\theta=\nu=1/2$ and in $4d$, $\theta=5/6$ and $\nu=5/12$ \cite{brydges}.)}
\end{center}
\end{table}

\begin{table}
\begin{center}
\begin{tabular}{|l|l|l|l|l|} \hline

              & $ST_1 (d=4)$     & $ST_2 (d=4)$  \\ \hline $\lambda$
&  $9.60\pm0.1$ & $10.2\pm0.1$ \\ \hline $\theta$      & $-0.13\pm0.1$
& $0.17\pm0.1$  \\ \hline $\nu_{pl}$    &  $0.33\pm0.02 $ &
$0.38\pm0.05$ \\ \hline $\nu_{\perp}$ &  $0.451\pm0.020$ &
$0.455\pm0.050$ \\ \hline
\end{tabular}
\vspace{0.5cm}
\caption{Estimates of critical exponents and growth constants from
Monte-Carlo simulations in four dimensions.}
\end{center}
\end{table}

\end{document}